\documentclass[12pt]{article} 
\usepackage[table,dvipsnames]{xcolor}
\usepackage[colorlinks, citecolor={blue}]{hyperref}
\usepackage{url}
\usepackage{amsfonts,amscd,amssymb}
\usepackage{amsthm,amsmath}
\usepackage{algorithm,algorithmicx,algpseudocode}
\usepackage{bm}
\usepackage{bbm} 
\let\oldtabular\tabular
\let\endoldtabular\endtabular
\renewenvironment{tabular}{\rowcolors{2}{white}{trevorblue!15}\oldtabular}{\endoldtabular}
\usepackage{verbatim}
\usepackage{graphicx}
\usepackage{setspace}
\usepackage{natbib}
\usepackage[margin=1in]{geometry}
\usepackage{enumitem}
\usepackage{listings}
\usepackage[textsize=tiny]{todonotes}
\usepackage{tikz}
\usepackage{etoolbox}
\usepackage{appendix}
\usepackage[format=plain,
font=it]{caption}
\usepackage{subcaption}
\usepackage{wrapfig}
\usepackage{xr}
\usepackage{booktabs}
\usepackage{multirow}
\usepackage{authblk}
\usepackage{adjustbox}
\usepackage{floatpag}

\allowdisplaybreaks

\newtoggle{quickdraw}
\toggletrue{quickdraw} 

\definecolor{lightgrey}{rgb}{0.9,0.9,0.9}
\definecolor{darkgreen}{rgb}{0,0.3,0}

\definecolorset{rgb}{}{}{darkred,0.8,0,0;darkgreen,0,0.5,0;darkblue,0,0,0.5}

\definecolor{trevorblue}{rgb}{0.330, 0.484, 0.828}
\definecolor{trevoryellow}{rgb}{0.829, 0.680, 0.306}

\usepackage{color}
\usepackage{fancyvrb}

\DefineVerbatimEnvironment{Highlighting}{Verbatim}{commandchars=\\\{\}}
\usepackage{framed}
\definecolor{shadecolor}{RGB}{241,243,245}

\newtheorem{example}{Example}

\DeclareMathOperator*{\argmin}{\arg\!\min}

\title{Scaling Hawkes Processes\footnote{Chapter 8 in Loeffler, C., Rosenberger, J.L., and Xue, L. (Eds.). Gun
		Violence: The Statistical Issues, (2026): Routledge.}}
\date{}

\author[1,2]{Seyoon Ko}
\author[2]{Jasen Zhang}
\author[2]{Andrew J.~Holbrook}
\affil[1]{Department of Mathematics, University of California, Los Angeles}
\affil[2]{Department of Biostatistics, University of California, Los Angeles}

\begin{document}

\maketitle

\begin{abstract}
Hawkes processes (HP) are a large class of stochastic point process models scientists have used to analyze contagion phenomena ranging from earthquakes, infectious diseases and biological neurons to financial trading activity, memes on social media and gun violence.  We introduce applications of HP to the latter before reviewing general strategies for fitting HP to data, paying attention to the influence of model structure on computational scalability considerations.  We then apply a recently developed high-performance computing powered Bayesian inference strategy for the spatiotemporal HP analysis of 412,376 acts of gun violence in the U.S.~between 2014 and 2024.  We finish with a discussion of model fit and directions for future research.
\end{abstract}


%

\newcommand{\Gen}{\mathcal{G}}
\newcommand{\x}{\mathbf{x}}
\newcommand{\X}{\mathbf{X}}
\newcommand{\dd}{\mbox{d}}
\newcommand{\Id}{\mathbf{I}}
\newcommand{\one}{\boldsymbol{1}}
\newcommand{\ttimes}{\mathbf{t}}
\newcommand{\bbeta}{\boldsymbol{\beta}}
\newcommand{\aalpha}{\boldsymbol{\alpha}}
\newcommand{\ttheta}{\boldsymbol{\theta}}
\newcommand{\mmu}{\boldsymbol{\mu}}
\newcommand{\z}{\mathbf{z}}

\section{Introduction}

HP model event contagion.  As such, this large class of models finds use in the analysis of earthquakes \citep{hawkes1973cluster,ogata1988statistical,zhuang2004analyzing}, viral memes \citep{yang2013mixture,mei2017neural}, biological neural activity \citep{linderman2014discovering,truccolo2016point,linderman2017bayesian}, viral epidemics \citep{kim2011spatio,meyer2014power,choi2015constructing,rizoiu2018sir,kelly2019real,holbrook2022bayesian,holbrook2022viral} and financial transactions \citep{embrechts2011multivariate,chavez2012high,hardiman2013critical,hawkes2018hawkes}.  In all these applications, the observation of an event (whatever it may be) at a specific time increases the probability of observing another event soon after.  

 In addition to dates or timestamps, individual data may contain spatial coordinates, categorical assignments or some other information about the events they describe.  Researchers have developed different kinds of HP to account for different kinds of data.  Applying an HP often involves fitting the model to data observed.  The exact method one should use to fit a specific HP model depends on the kind of model, the size of the data, the purpose of the analysis, the computational resources at hand and the amount of time one would require to implement the method.  A streamlined introduction, Section \ref{sec:examp} provides examples involving five methods for fitting HP of various forms, paying attention to each method's time complexity as a function of data count and relevant dimensions.
Such computational considerations are paramount in the age of big data, especially since many different approaches to fitting various HP to $N$ observed events typically  require a quadratic $O(N^2)$ number of floating point operations.  Such computational burdens prevail  regardless of data application, but scientists working in fields featuring massive quantities of events must implement their HP models with scalability in mind. 

As a primary example, HPs have  emerged as statistical tools enabling the use of data to mathematically characterize the contagiousness of violence, a premise advanced within the criminology literature \citep{loftin1986assaultive,Fagan_Wilkinson_Davies_2007,taylor2013contagion}.   \citet{mohler2014marked} develops a novel spatiotemporal HP (StHP) model for homicides conditioned on other violent crimes.  \citet{mohler2014marked} analyzes subsets of 80,000 events in Chicago from 2007 to 2012 using an EM algorithm (cf., Section \ref{sec:examp}, Example \ref{ex:em}),  requiring roughly 200 iterations for an analysis involving 3-year subsets of the data.  For each subsets, homicides are on the order of thousands while total violent crimes are in the tens-of-thousands; across all EM iterations the total number of floating point operations is $O(10^{9})$, reportedly requiring one hour of time on a personal laptop.  \citet{linderman2014discovering} use the multivariate HP (MHP) framework described in  Section \ref{sec:examp}, Example \ref{ex:scott}, to analyze over 1,600 gang-related homicides across 77 Chicago neighborhoods from 1980 to 1995.  The authors' careful GPU implementation achieves 5-50 Gibbs updates per second across all examples considered, and it is not clear exactly where their Chicago gang-violence analysis lands on this interval.  The authors do not seem to report the number of Gibbs updates required, but it is reasonable to assume that they use orders-of-magnitude more than the number of EM updates required to fit the albeit smaller model of \citet{mohler2014marked}.

\citet{loeffler2018gun} use an StHP to analyze over 9,000 gunshots recorded by an acoustic gunshot locator system (AGLS) in Washington D.C.~from 2010 to 2012.  After specifying prior distributions on StHP parameters, \citet{loeffler2018gun} perform Bayesian inference using Hamiltonian Monte Carlo with the help of the  Stan probabilistic programming language.  In total, the authors simulate four MCMC chains for 200 iterations each within an unspecified amount of time.  For convergence diagnostics, they do not report effective sample sizes (ESS) but do report a Gelman-Rubin $\hat{R}$ statistic close to 1.  Again working within the Bayesian framework, \citet{holbrook2021scalable} scale a similar StHP model to over 85,000 gunshots recorded by the same AGLS in Washington D.C.~from 2006 to 2019 with the help of hand-coded GPU and multi-core CPU routines for the StHP likelihood.  The need for such parallelization arises from the fact that a 10-fold increase in events results in a 100-fold increase in likelihood calculation floating point operations.   \citet{holbrook2021scalable} implement their fast likelihood calculations within an adaptive Metropolis-Hastings (MH) algorithm \citep{metropolis1953equation,hastings1970monte}, generating 4 Markov chain Monte Carlo (MCMC) chains of length 10,000, each requiring 10 hours and obtaining a minimum ESS of 1,700.  Using the same fast likelihood calculations within a similar adaptive MH scheme, \citet{holbrook2022bayesian} extend this model to include event-specific latent locations with prior probabilities reflecting the spatial precision of location measurements. Although such a data augmentation strategy improves model fit in simulations, the resulting $O(N)$ increase in model parameters restricts applicability to events numbered in the low thousands.

Recently, \citet{ko2024scaling} introduce a multi-GPU, high-performance computing strategy for fitting Bayesian StHP models to big spatiotemporal data using the adaptive MH algorithms at the core of \citet{holbrook2021scalable,holbrook2022bayesian}.   \citet{ko2024scaling} apply their computational inference framework to the analysis of roughly one million COVID-19 cases in the U.S.; in Section \ref{sec:meth}, we show that this framework readily translates to gun violence data.  This work is mainly computational in nature, and we do not discuss other contemporaneous contributions in this area that improve the interpretability of HP analyses, expand the expressivity of HP models or enable more effective forecasting from HP models once fit \citep{sha2020interpretable,park2021investigating,brantingham2021recent,zhu2022spatiotemporal,dong2023spatiotemporal}.  We do, however, note that the general fine-grained parallelization strategies for HP likelihood calculations are largely applicable to many of these works and hope that the computational contributions we consider serve as a road map for scaling ever more interpretable, flexible and complex HP models to big data.  Before entering in on the analysis of big American gun violence data, we review several types of HP and the computational complexity of  strategies for fitting them.

\section{Strategies for Fitting Hawkes Processes}\label{sec:examp}

An HP is a stochastic point process model that exhibits self-exciting behavior \citep{hawkes1971point,hawkes1971spectra,hawkes1972spectra}.  In its basic form, an HP characterizes  collections of random events in time $\{t_1,t_2 ,\dots\}\in [0,\infty)$ using a conditional intensity function
\begin{align}\label{eq:int}
\lambda^*(t)=\lambda(t|\, t_n<t) = \lambda_0 + \sum_{t_n<t} \xi(t-t_n)
\end{align}
that describes the instantaneous rate of event arrival at time $t$ given preceding event times. Here, $\lambda_0>0$ is a fixed background rate and $\xi(\cdot)>0$ is the excitation function or triggering kernel, the positivity of which increases $\lambda^*(t)$ immediately after each event $t_n$.  This increase in the arrival rate may correspond to the occurrence of more events, each of which provides its own contribution to $\lambda^*(t)$.   \citet{hawkes1974cluster} relate the conditional intensity-based representation \eqref{eq:int}  to an immigrant-offspring process representation that starts with a generation of immigrant events 
$ \Gen_0 = \{t_{0,1}, t_{0,2}, \dots\}$ following a homogeneous Poisson process with rate $\lambda_0$.  Each background event $t_{0,n}$ gives birth  to $O_{1,n}\sim \mbox{Poisson}(\beta)$ offspring events, where $\beta$ is the branching number 
\begin{align}\label{eq:branching}
	\beta = \int_{t_{0,n}}^{\infty}  \xi(t-t_{0,n}) dt = \int_{0}^{\infty}  \xi(t) dt \, .
\end{align}
If $O_{1,n}>0$, the $O_{1,n}$ offspring occur at times  $\Gen_{1,n} = \{t_{1,n,1}, \dots t_{1,n,O_{1,n}}\} \stackrel{iid}{\sim} \xi(t-t_{0,n})/\beta$.  The second generation is the union $\Gen_1 = \cup_{n \in \Gen_0} \Gen_{1,n} = \{t_{1,1} \dots t_{1,O_1}\}$, where $O_1=\sum_n O_{1,n}$ and it is convenient to drop the index relating a child to its parent. The process continues in perpetuity, with each generation $\Gen_{g+1}$ consisting of offspring from members of the previous generation $\Gen_{g}$ and the entire HP sample path being the union $\cup_{g} \Gen_{g}$.  

Myriad extensions to the basic HP exist and take forms adapted to the modeling of data  arising from various phenomena.  Both the conditional intensity and the immigrant-offspring representation continue to apply for many of these modeling extensions.  As is common in many areas of statistics, one may fit an HP model to observed data using the method of moments or likelihood-informed methods such as maximum likelihood, EM or Bayesian inference.  Depending on model specifics and the technique one uses to fit one's model, it may be more convenient (computationally or otherwise) to use one of the HP representations or the other.  This section is not a comprehensive review of the HP zoo and its applications.  The reader may consult \citet{laub2015hawkes,laub2021elements} for more comprehensive reviews, \citet{hawkes2018hawkes} for a review of the HP in finance and \citet{reinhart2018review} for a review of the spatiotemporal HP.  Here, we quickly touch on a small number of HP models and briefly discuss a few methods for fitting these models, paying attention to methods' relative benefits and the quadratic $O(N^2)$ time-complexity required to fit most interesting HP models to $N$ observations, regardless of representation used.   

\begin{example}[Method of Moments]
 	\citet{da2014hawkes} consider a basic exponential HP with rate function
	\begin{align}\label{eq:exp1}
		\lambda^*(t) = \lambda  + (\lambda_0-\lambda) e^{-\omega t}  + \xi_0 \sum_{t_n<t}  e^{-\omega(t-t_n)} 
	\end{align}
and derive the asymptotic moments for the number of jumps over intervals of length $\tau$. Letting $N_t$ be the number of events on the interval $[0,t]$, these moments include
\begin{align}
	\mu_{\tau,1} &= \lim_{t\rightarrow \infty} \mathbb{E} (N_{t+\tau}-N_t | \lambda_0) = \frac{\lambda \omega}{\omega-\xi_0} \tau \\
		\mu_{\tau,2} &= \lim_{t\rightarrow \infty} \mathbb{E}\left( (N_{t+\tau}-N_t)^2 | \lambda_0\right) - \mu_{\tau,1}^2  =  \frac{\lambda \omega}{\omega-\xi_0} \left(\frac{ \tau \omega^2}{(\omega-\xi_0)^2}   + \left( 1-  \frac{\omega^2}{(\omega-\xi_0)^2} \right)  \frac{1- e^{(\xi_0-\omega)\tau}}{\omega-\xi_0}    \right) \nonumber
\end{align}
as well as the autocovariance between counts of intervals separated by arbitrary lags.  One may obtain corresponding empirical moments $\hat{\mu}_{\tau,1}$ and $\hat{\mu}_{\tau,2}$ by dividing the data observation interval $[t_0,T]$ into $B$ bins of length $\tau$ and summing over bin-wise event counts $C_b$:
\begin{align}
	\hat{\mu}_{\tau,1} = \frac{1}{B} \sum_{b=1}^B C_b \, , \quad 	\hat{\mu}_{\tau,2} = \frac{1}{B} \sum_{b=1}^B C_b^2 -  \hat{\mu}_{\tau,1}^2 \, .
\end{align}
Here,  one may specify $t_0$ to discard a certain number of early events as burn-in while ensuring that $\tau$ divides the observation interval.   \citet{da2014hawkes} use the generalized method of moments \citep{hall2005generalized} to perform moment matching by defining $\ttheta=(\lambda,\xi_0,\omega)$, letting $\hat{\mmu}$ and $\mmu(\cdot)$ be vectors of emprical and theoretical moments, respectively, and using Levenberg-Marquardt \citep{levenberg1944method,marquardt1963algorithm} to find
\begin{align*}
	\hat{\ttheta} = \argmin_{\ttheta}  \left(\hat{\mmu}_{\tau}-\mmu_{\tau}(\ttheta)\right)^T\left(\hat{\mmu}_{\tau}-\mmu_{\tau}(\ttheta)\right)\, .
\end{align*}
In fact, any nonlinear optimization routine will do, and the generalized method of moments provides consistent estimators that satisfy asymptotic normality.  In terms of empirical estimator accuracy and variability, \citet{da2014hawkes}  show that their method can perform better or worse than maximum likelihood (discussed below) depending on the moments used.  They do not communicate speedups over the latter but note that computing for their method is ``negligible'' and that even the fast maximum likelihood method of \citet{ozaki1979maximum} ``takes a few minutes.''
\end{example}

A  drawback for the method of moments is that one must know  theoretical moments in closed form.  While this may be possible for simple models such as the exponential HP, the method of moments does not provide a recipe for scalably robust data science involving repeated application of multiple competing models within a single scientific workflow.  What it gains in its comparably fast $O(N)$ time-complexity, it loses in person-hours since each new model requires erudite derivations, examples of which are provided in the  impressive work of \citet{daw2023matrix}.   On the other hand, maximum likelihood provides a more widely applicable method for fitting HP but generally requires $O(N^2)$  floating-point operations for each gradient evaluation, with \citet{ozaki1979maximum} providing one prominent exception.

\begin{example}[Maximum Likelihood]\label{ex:ml}
For a point process with conditional intensity $\lambda^*(t)$ parameterized by $\ttheta$, the likelihood of observing events  $\{t_1 ,\dots, t_N\}$ on an observation interval $[0,T]$ is \citep{rubin1972regular,daley2003conditional}
\begin{align}\label{eq:basic_lik}
	L( \{t_n\}_{n=1}^N | \ttheta, T) =    e^{-\int_0^T \lambda^*(t)dt} \prod_{n=1}^N \lambda^*(t_n) =e^{-\Lambda(T)} \prod_{n=1}^N \lambda^*(t_n) \, ,
\end{align}
and the log-likelihood for the HP with conditional intensity \eqref{eq:int} is
\begin{align}\label{eq:basic_ll}
	\ell( \{t_n\}_{n=1}^N | \ttheta, T) =  -\Lambda(T)+  \sum_{n=1}^N  \log \left( \lambda_0 + \xi_0 \sum_{t_{n'}<t_n} \xi(t_n-t_{n'}) \right)  .
\end{align}
	For the simplified version of the exponential HP condidtional intensity  \eqref{eq:exp1}
	\begin{align}
		\lambda^*(t) = \lambda_0  + \xi_0 \sum_{t_n<t}  e^{-\omega(t-t_n)}  \, ,
	\end{align}
the log-likelihood is
    \begin{align}\label{eq:lik_naive}
	\ell( \{t_n\}_{n=1}^N | \ttheta, t_N)= -\lambda_0 t_N + \sum_{n=1}^N \left( \log \left(\lambda_0 + \xi_0 \sum_{n'=1}^{n-1} e^{-\omega(t_n-t_{n'})}  \right) + \frac{\xi_0}{\omega}\left(  e^{-\omega(t_{N}-t_n)} -1 \right) \right) ,
\end{align}
where we follow common practice and take $t_N$ as endpoint of the observation interval. 
While \eqref{eq:lik_naive} and its double summation ostensibly suffer $O(N^2)$ time-complexity, a recursion reveals linear complexity in $N$ \citep{ozaki1979maximum,laub2015hawkes}.  For $n=2,\dots,N$, let $\psi(n) = \sum_{n'=1}^{n-1}e^{-\omega(t_n-t_{n'})}$. Then
\begin{align*}
	\psi(n) &= e^{-\omega t_n + \omega t_{n-1}}\sum_{n'=1}^{n-1} e^{-\omega t_{n-1} + \omega t_{n'}} = e^{-\omega (t_n- t_{n-1})} \left(1+ \sum_{n'=1}^{n-2} e^{-\omega t_{n-1} - \omega t_{n'})}\right) \\
	&=e^{-\omega (t_n - t_{n-1})} \left(1+\psi(n-1) \right)\, .
\end{align*} 
Letting $\psi(1)=0$, the log-likelihood is
\begin{align*}
		\ell( \{t_n\}_{n=1}^N | \ttheta, t_N)= -\lambda_0 t_N + \sum_{n=1}^N \left( \log \left(\lambda_0 + \xi_0 \psi(n)  \right) + \frac{\xi_0}{\omega}\left(  e^{-\omega(t_{N}-t_n)} -1 \right) \right)  .
\end{align*}
Through similar recursions, the log-likelihood gradient and Hessian yield $O(N)$-time forms that one may implement within one's favorite nonlinear optimization algorithm.
\end{example}

Likelihood-based methods flexibly adapt to HP model extensions but pay the price in their $O(N^2)$ time-complexity.  Log-likelihood formula \eqref{eq:basic_ll} continues to apply if one specifies the power-law kernel $	\xi(t) = \xi_0 \omega / (1+\omega t)^{p+1}$, but no known recursion facilitates similar $O(N)$ computations.  
Beyond simple changes to $\xi(\cdot)$, likelihoods are available for more complex processes. A basic spatiotemporal HP (StHP) models the usual temporal event data $t_n$ along with corresponding  locations $\x_n$ within some spatial domain $\mathcal{X}$ using the conditional intensity
\begin{align}
	\lambda^*(t,\x) = \lambda_0(\x) + \sum_{t_n<t} \xi(t-t_n, \x-\x_n) 
\end{align}
for $\lambda_0(\cdot),\xi(\cdot,\cdot)\geq0$, and the likelihood for a general StHP takes the recognizable form  \citep{daley2003conditional}
\begin{align}\label{eq:sthplik}
L( \{(t_n,\x_n)\}_{n=1}^N | \ttheta, T) =    e^{-\int_{\mathcal{X}}\int_0^T \lambda^*(t,\x)dtd\x} \prod_{n=1}^N \lambda^*(t_n,\x_n) = e^{-\Lambda(T,\mathcal{X})} \prod_{n=1}^N \lambda^*(t_n,\x_n) \, .
\end{align}
Unfortunately, \citet{veen2008estimation} show that the log-likelihood can be relatively flat around its maximum for some important StHP and that this leads to convergence issues for popular numerical optimization algorithms such as Newton-Raphson.

\begin{example}[Expectation Maximization]\label{ex:em}
	Relying on the immigrant-offspring representation of the HP, \citet{veen2008estimation} propose augmenting each data point $(t_n,\x_n)$ for $n \in \{1,\dots,N\}$ with an unobserved latent variable $z_n \in \{0,1,\dots,n-1\}$ that indicates the parent of event $n$ ($z_n=0$ for $n$ an immigrant event) and using the EM algorithm \citep{dempster1977maximum} to maximize the  likelihood
	 \begin{align}
	 	L\left( \{(t_n,\x_n)\}_{n=1}^N | \ttheta, T\right) = \sum_{z_1,\dots,z_N} L\left(\{(t_n,\x_n, z_n)\}_{n=1}^N | \ttheta, T\right)   
	 \end{align}
 obtained by marginalizing the complete-data likelihood
 \begin{align*}
L\left(\{(t_n,\x_n. z_n)\}_{n=1}^N | \ttheta, T\right) \propto \prod_{n=1}^N  \lambda_0  (\x_n)^{\mathcal{I}_{[z_n=0]} } \prod_{n'=1}^{n-1} \xi(t_n-t_{n'}, \x_n-\x_{n'}) ^{\mathcal{I}_{[z_n=n']}}  \, ,
 \end{align*}
where $\mathcal{I}_{[\cdot]}$ is the indicator function.  The EM algorithm of \citet{veen2008estimation} begins with  some parameter value $\ttheta^{(0)}$ and alternates between expectation and maximization steps, performance of which together contribute to a single iteration.  At iteration $s$, the  E-step uses parameter $\ttheta^{(s-1)}$  to obtain
\begin{align}
	\mbox{\emph{Pr}}(z_n=n')= \mathbb{E}_{z_n}(\mathcal{I}_{[z_n=n']}) = \begin{cases} 
		\frac{\xi(t_n-t_{n'},\x_n-\x_{n'})}{\lambda^*(t_n,\x_n)} & n' \in \{1,\dots,n-1\} \\
		\frac{\lambda_0(\x_n)}{\lambda^*(t_n,\x_n)} & n'=0
	\end{cases} 
\end{align}
for $n\in \{1,\dots,N\}$.  The M-step then maximizes the expected complete-data log-likelihood
\begin{align}\label{eq:Q}
	\mathbb{E}_{z_1,\dots,z_N} \ell\left(\{(t_n,\x_n)\}_{n=1}^N | \ttheta, T,\{z_n\}_{n=1}^N\right) &\propto  \sum_{n=1}^N 	\mbox{\emph{Pr}}(z_n=0) \log \lambda_0(\x_n) \\ \nonumber
	+& \sum_{n=1}^{N}   \sum_{n'=1}^{n-1}  \mbox{\emph{Pr}}(z_n=n') \log  \xi(t_n-t_{n'}, \x_n-\x_{n'}) \, .
\end{align}
\citet{veen2008estimation} perform this step by setting the gradient of \eqref{eq:Q} with respect to $\ttheta$ to $\boldsymbol{0}$ and solving the resulting nonlinear equations. The algorithm continues until  $\lVert\ttheta^{(s+1)}- \ttheta^{(s)}\rVert$ is less than some prespecified small number. While the algorithm confers stability and robustness compared to numerical optimization approaches, it does not overcome the $O(N^2)$ time-complexity facing likelihood-based methods: its $\binom{N}{2}$ probabilities $\mbox{\emph{Pr}}(z_n=n')$   require $O(N^2)$ floating-point operations.  \citet{veen2008estimation} successfully apply their algorithm to the analysis of 6,796 earthquakes in Southern California but do not communicate the amount of time required to fit their model.

\end{example}

This strategy solves the  aforementioned convergence issues suffered by numerical optimization techniques for finding the maximum likelihood estimator insofar as the EM algorithm provably moves uphill at every iteration, attaining a local maximum upon convergence.  Despite the fact that the EM algorithm often results in a consistent estimator \citep{dempster1977maximum}, the lack of a global maximization guarantee complicates uncertainty quantification.  There is no general guarantee of asymptotic normality that one would otherwise use to construct confidence intervals  for the EM algorithm's estimator.   Within the Bayesian paradigm, one may use the same data augmentation strategy as \citet{veen2008estimation} to implement inference algorithms that provide uncertainty quantification with asymptotic exactness guarantees. Example \ref{ex:scott} provides one such instance.

The multivariate Hawkes Process (MHP) extends the basic HP by introducing marked events $\{t_n, \gamma_n\}_{n=1}^N$ for $\gamma_n \in \{1, \ldots, K\}$. MHPs categorize events into one of $K$ classes with $K$ corresponding intensity functions $\{\lambda^*_k(t)\}_{k=1}^K$. For any time $t$, preceding events influence the intensity function in a manner analogous to the basic HP \eqref{eq:int}:
\begin{align}\label{eq:MHP}
	\lambda_k^*(t) &= \lambda_{0, k}(t) +  \sum_{t_n < t} \xi_{\gamma_n, k}(t - t_n).
\end{align}
Here, $\lambda_{0, k}(t)$ is the background rate of class $k$, and  $\xi_{k', k}(\cdot)$ is the excitation function that describes the contribution of events of class $k'$ to the intensity of class $k$. The MHP likelihood takes a form analogous to \eqref{eq:basic_lik}:
\begin{align}\label{eq:MHP_Marginal}
	& L\big(\{t_n, \gamma_n\}_{n=1}^N  \mid \{\lambda_{0, k}(t)\}_{k=1}^K, \{\xi_{k', k}(t)\}_{k',k = 1}^K, T\big) = e^{- \Lambda(T, K)} \prod_{n=1}^N  \lambda_{\gamma_n}^*(t_n),
\end{align}
where $\Lambda(T, K) = \int_0^T \sum_{k=1}^K \lambda_k^*(t) dt$. Its computation has $O(N(N+K))$ time-complexity. 
\begin{example}[Markov Chain Monte Carlo]\label{ex:scott} Within a Bayesian hierarchical model formulation,	\citet{linderman2014discovering} build an MHP with an immigration representation of the likelihood in \eqref{eq:MHP_Marginal}:
	\begin{align}\label{eq:MHP_Augment}
		 L\Big(\left \lbrace (t_n, \gamma_n, z_n) \right \rbrace_{n=1}^N \mid & \{\lambda_{0, k}(t)\}_{k=1}^K, \{\xi_{k', k}(t)\}_{k'{,} k = 1}^K, T\Big) \nonumber  \\
		&= e^{- \Lambda(T, K)}\prod_{n=1}^N   \lambda_{0, \gamma_n}(t_n)^{\mathcal{I}_{[z_n=0]}} \; \xi_{z_n, \gamma_n} (t_n-t_{z_n})^{\mathcal{I}_{[z_n\neq 0]}}.
	\end{align}
	Just as in Example \ref{ex:em}, $\{z_n\}_{n=1}^N$ represent the parents of events $\{t_n\}_{n=1}^N$, with $z_n=0$ denoting that event $n$ is an immigrant event. The authors also choose the following parameterization of the baseline intensity and excitation function
\begin{align*}
	\lambda_{0, k}(t) = \mu_k + \alpha_k e^{y(t)}, \quad y(t) \sim \mathcal{GP}(0, \kappa(t, t')), \quad \xi_{k,k'}(t) = A_{k, k'} W_{k, k'} g_{\boldsymbol{\theta}_{k, k'}}(t).
\end{align*}
A Gaussian process $y(t)$ \citep{rasmussenwilliamsgp} and constants $\mu_k$ and $\alpha_k$ specify the background rate, while a binary adjacency matrix $\mathbf{A} \in \{0, 1\}^{K \times K}$, a matrix of non-negative weights $\mathbf{W} \in \mathbb{R}_+^{K \times K}$, and a logistic-normal density function $g_{\boldsymbol{\theta}_{k', k}}(t)$ with parameters $\{\boldsymbol{\theta}_{k', k}\}_{k', k=1}^K$ characterize the excitation function. 
	
\citet{linderman2014discovering} develop a novel Gibbs sampler  \citep{geman_gibbs, gelfand90a} to generate samples from the posterior distribution of the MHP parameters. Gibbs sampling offers a robust procedure for posterior inference by iteratively sampling subsets of parameters from their corresponding conditional posterior distributions. The authors sample the conditional posteriors of $\mathbf{W}$ and  $\{\boldsymbol{\theta}_{k, k'}\}_{k, k' = 1}^K$ with specified conjugate priors and $y(t)$ (evaluated on a grid) with elliptical slice sampling \citep{murray2010elliptical}. The authors specify Aldous-Hoover graph priors \citep{lloyd_2012} on $\mathbf{A}$ to render the elements of $\mathbf{A}$ conditionally independent, but the augmented MHP likelihood \eqref{eq:MHP_Augment} induces dependencies between the rows of $\mathbf{A}$. Nonetheless, the columns of $\mathbf{A}$ are still independent, enabling the parallelization of Gibbs updates. The authors use collapsed Metropolis-within-Gibbs to sample the columns of $\mathbf{A}$ with the marginalized likelihood \eqref{eq:MHP_Marginal}. Finally, each parent variable $z_n$ updates from a discrete distribution defined over the set $\{0, 1, \ldots, n-1\}$, and all $N$ parent variables are conditionally independent. 

\citet{linderman2014discovering} address major computational bottlenecks by pre-specifying the Gaussian process kernel parameters to avoid repeated $O(N^3)$ computations. Additionally, they use parallelization to exploit the conditional independence between the columns of $\mathbf{A}$ and elements of $\{z_n\}_{n=1}^N$ within their respective blocks. Through parallelization, they reduce the effective time-complexity of updating $\{z_n\}_{n=1}^N$ from $O(N^2)$ to $O(N)$, and updating the columns of $\mathbf{A}$ from $O(N^2K)$ to $O(N^2)$. Furthermore, assuming the bounded support of a logistic-normal density reduces the time-complexity of updating $\mathbf{A}$ by a significant constant. Finally, parallelized binary reductions summing $N$ contributions to a likelihood reduce $O(N)$ to $O(\log{N})$ time-complexity \citep[cf.,][]{holbrook2021massive}. 

\citet{linderman2014discovering} mention that their parallelized Gibbs sampler achieves 5 to 50 iterations per second for the examples they consider, but they do not report the number of iterations required. The authors implement their algorithm on S\&P 100 index dataset collected at $1$s intervals during the week from September 28 through October 2, 2009. Events denote times when a stock price changes by $\pm 0.1\%$ of its current price. The dataset has $K = 100$ processes and $N = 182{,}037$ events. They also model gang-related homicides between 1980 and 1995 in Chicago with $K = 77$ communities and $N = 1{,}637$ homicide events. 

\end{example}
The preceeding HPs only allow event occurrences to increase the probability of future events. Nonlinear HPs expand the modeling flexibility of linear HPs with a nonlinear transfer function $f: \mathbb{R} \rightarrow \mathbb{R}_+$ that allows one to drop the non-negativity assumption on the excitation function $\xi(\cdot)$ within the conditional intensity function:
\begin{align}\label{eq:nonlinear_hp}
    \lambda^*(t)= f\left(\lambda_0 + \sum_{t_n<t} \xi(t-t_n)\right).
\end{align}
The inclusion of a nonlinear transfer function allows nonlinear HPs to model inhibitory events and non-additive contributions of past events \citep{nonlinear_hawkes}. Recent additions to the nonlinear HP literature focus on using deep neural networks to model nonlinear conditional intensities such as \eqref{eq:nonlinear_hp}.
\begin{example}[Stochastic Gradient Descent]\label{ex:neural}
	\citet{mei2017neural} introduce neural Hawkes processes (NHPs), which they describe as ``neurally self-modulating MHPs" that have the ability to model nonlinearities in their conditional intensity functions. For all times $t > 0$ and all classes $k \in \{1, \ldots, K\}$, the class-specific conditional intensity function is
\begin{equation*}
    \lambda_k^*(t) = f_k(\mathbf{w}_k^{\top} \mathbf{h}(t))\, ,
\end{equation*}
where $f_k(\cdot)$ is the softplus function, $\mathbf{w}_k \in \mathbb{R}^D$ is a vector of weights for class $k$, and $\mathbf{h}(t) \in (-1, 1)^{D}$ is a hidden state vector at time $t$. The authors model NHPs using a continuous-time long-short term memory (LSTM) architecture, inspired by the familiar discrete-time LSTM \citep{Hochreiter_1997}. In the continuous-time framework, long-term memory  $\mathbf{c}(t)$ and short-term memory $\mathbf{h}(t)$ are $D$-dimensional vectors that update in the continuous interval $(t_n, t_{n+1}]$ between events $n$ and $n+1$:
\begin{align*}
	\mathbf{c}(t) = \overline{\mathbf{c}}_{n+1} + (\mathbf{c}_{n+1} - \overline{\mathbf{c}}_{n+1}) e^{\left(- \boldsymbol{\delta}_{n+1}(t - t_n)\right)}, \quad \mathbf{h}(t) = \mathbf{o}_n \odot (2 \sigma(2 \mathbf{c}(t)) - 1) \, ,
\end{align*}
where $\sigma(\cdot)$ is the elementwise sigmoid function and $\odot$ is the Hadamard product. Here, long-term memory $\mathbf{c}(t)$ exponentially decays at rate $\boldsymbol{\delta}_{n+1}$ towards a steady-state value $\overline{\mathbf{c}}_{n+1}$. Other LSTM values update instantaneously at time $t_{n+1}$ with update rules
\begin{align*}
    \mathbf{i}_{n+1} & \leftarrow \sigma\left(\mathbf{W}_i \mathbf{k}_n + \mathbf{U}_i \mathbf{h}(t_n) + \mathbf{d}_i\right) & \mathbf{\overline{i}}_{n+1} & \leftarrow \sigma\left(\mathbf{W}_{\overline{i}} \mathbf{k}_n + \mathbf{U}_{\overline{i}} \mathbf{h}(t_n) + \mathbf{d}_{\overline{i}}\right)\\
    \mathbf{f}_{n+1} & \leftarrow \sigma\left(\mathbf{W}_f \mathbf{k}_n + \mathbf{U}_f \mathbf{h}(t_n) + \mathbf{d}_f\right) & \mathbf{\overline{f}}_{n+1} & \leftarrow \sigma\left(\mathbf{W}_{\overline{f}} \mathbf{k}_n + \mathbf{U}_{\overline{f}} \mathbf{h}(t_n) + \mathbf{d}_{\overline{f}}\right)\\
    \mathbf{z}_{n+1} & \leftarrow 2 \sigma\left(\mathbf{W}_z \mathbf{k}_n + \mathbf{U}_z \mathbf{h}(t_n) + \mathbf{d}_z\right) - 1 & \mathbf{c}_{n+1} & \leftarrow \mathbf{f}_{n+1} \odot \mathbf{c}(t_n) + \mathbf{i}_{n+1} \odot \mathbf{z}_{n+1}\\
    \mathbf{o}_{n+1} & \leftarrow \sigma\left(\mathbf{W}_o \mathbf{k}_n + \mathbf{U}_o \mathbf{h}(t_n) + \mathbf{d}_o\right) & \overline{\mathbf{c}}_{n+1} & \leftarrow \overline{\mathbf{f}}_{n+1} \odot \overline{\mathbf{c}}_n + \overline{\mathbf{i}}_{n+1} \odot \mathbf{z}_{n+1}\\
    \boldsymbol{\delta}_{n+1} & \leftarrow \zeta\left(\mathbf{W}_d \mathbf{k}_n + \mathbf{U}_d \mathbf{h}(t_n) + \mathbf{d}_d\right).
\end{align*}
for nonlinear function $\zeta: \mathbb{R} \rightarrow \mathbb{R}_+$. The vector $\mathbf{k}_n \in \{0, 1\}^K$ is the one-hot encoding of the class of event $n$, and $\{\mathbf{W}_\ell \in \mathbb{R}^{D \times K},\mathbf{U}_\ell \in \mathbb{R}^{D \times D}, \mathbf{d}_\ell \in \mathbb{R}^D\}$ are LSTM parameters for $\ell \in \{i,f,z,o,d, \overline{i}, \overline{f}\}$.  

\citet{mei2017neural} use stochastic gradient descent (SGD) \citep{bottou2010large} to fit the LSTM parameters by minimizing the negative log-likelihood (cross-entropy loss)
\begin{align*}
	&- \ell\big(\{t_n, \gamma_n\}_{n=1}^N  \mid \{\lambda_{0, k}(t)\}_{k=1}^K, \{\xi_{k', k}(t)\}_{k',k = 1}^K, T\big) =  \Lambda(T, K) - \sum_{n=1}^N  \log{\lambda_{\gamma_n}^*(t_n)}
\end{align*}
arising from \eqref{eq:MHP_Marginal}.
In each iteration of SGD, the authors use Monte Carlo simulations to estimate the gradient of the cumulative intensity, $\nabla \Lambda(T, K)$: 
\begin{align}\label{eq:MC_estimate}
	\nabla \widehat{\Lambda}(T, K) = \frac{T}{M} \sum_{m=1}^M  \sum_{k=1}^K \nabla \lambda^*_k(t_m), \quad  t_1, t_2, \ldots, t_M \overset{iid}{\sim} \mbox{Uniform}  (0, T) \, .
\end{align}
The estimate $\nabla\widehat{\Lambda}(T, K)$ is an unbiased estimator of $\nabla \Lambda(T, K)$ since it is the sum of $MK$ true gradients. The computational bottleneck of SGD lies in the computation of $\widehat{\Lambda}$ with a time-complexity of $O(MND(D+K))$, where $D$ is the dimensionality of the LSTM memory. Computing $\nabla\lambda_k^*(t_m)$ in \eqref{eq:MC_estimate} requires forward-pass through $O(N)$ LSTM cells, each of which performs two matrix multiplication steps with $O(D^2)$ and $O(DK)$ time-complexity.  \citet{mei2017neural} fit their model with Theano, a now-defunct Python library built to enable massively parallel GPU-based acceleration of key backpropagation computations. The authors fit their model on a Twitter dataset \citep{zhao_tweets_2015} $(K = 3, N = 166{,}076)$ and the MemeTrack dataset \citep{snapnets} $(K = 5{,}000, N = 1{,}500{,}000)$, but they do not mention the time required to do so.

\end{example}

\section{Analysis of American Gun Violence Data}\label{sec:meth}

 \begin{figure}[!t]
 	\centering
 	\includegraphics[width=1\textwidth]{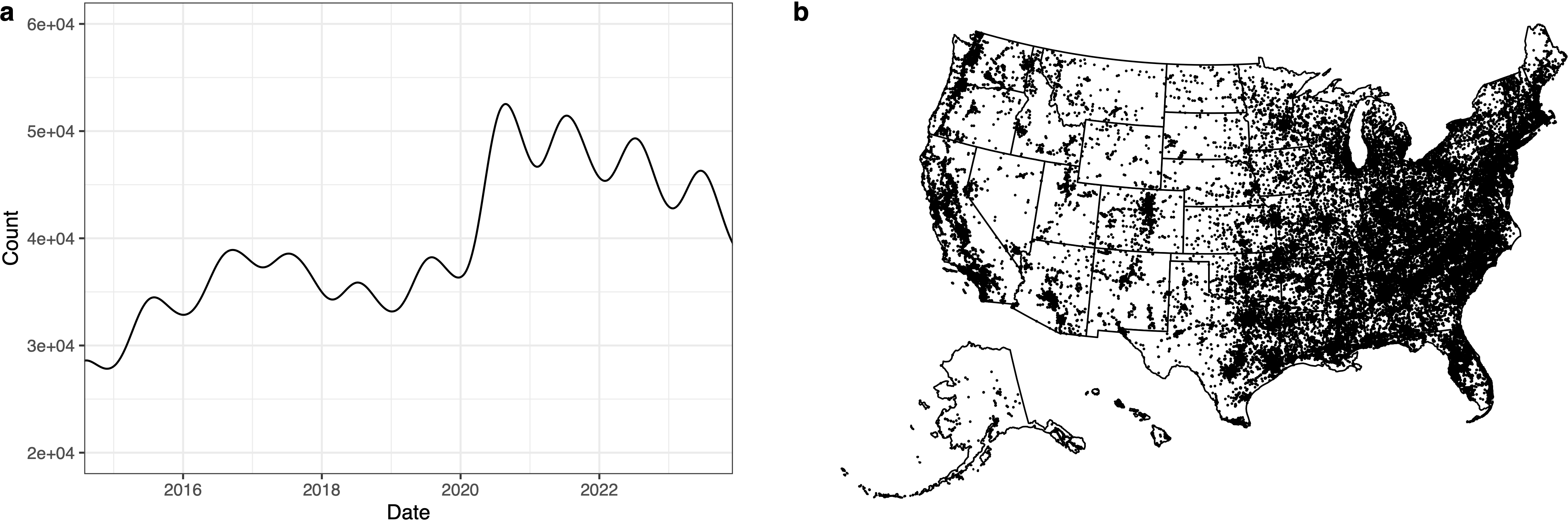}
 	\caption{Temporal (\textbf{a}) and spatial (\textbf{b}) distributions of 412,376 American gun violence events recorded from 2014 to 2024 and provided by the Gun Violence Archive.  We use the high-performance computational inference strategy of \citet{ko2024scaling} to perform Bayesian inference and analyze this data in only 10 hours despite the task requiring more than O($16 \times 10^{10}$) floating-point operations per iteration of Markov chain Monte Carlo.}\label{fig:violence}
 \end{figure}

 In this section, we use data from the American gun violence epidemic to demonstrate how high-performance computing (HPC) can help overcome the HP's usual quadratic time-complexity to  enable HP-based analyses of large-scale public health data.
We model spatiotemporal data $\{(\x_n,t_n)\}_{n=1}^N$ similar to that of Example \ref{ex:em} using an StHP with conditional intensity function
\begin{align}\label{eq:condInt}
\lambda^*(t,\x)= \frac{\lambda_0}{|\mathcal{X}|} \sum_{n=1}^N \mathcal{I}_{[t\neq t_n]}  \phi_1 (t|t_n, \tau_t) + \frac{\xi_0}{\omega_t} \sum_{t_n<t}  e^{-  (t-t_n)/ \omega_t }  \phi_2( \x | \x_n,\omega_x) \, ,
\end{align}
with $\mathcal{I}$ the indicator function, $|\mathcal{X}|$ the area of the spatial domain, $\phi_1$ the $1$-dimensional Gaussian density, $\phi_2$ the $2$-dimensional spherical Gaussian density, and $\lambda_0$, $\tau_t$, $\omega_x$, $\omega_t$ and $\xi_0$ positive parameters.  Here, $\tau_t$ is the background temporal lengthscale, $\omega_t$ is the triggering kernel's temporal lengthscale, and $\omega_x$ is the triggering kernel's spatial lengthscale. 
We use the same prior specification for these parameters as that of \citet{ko2024scaling} but note that priors appear to have little influence in the big data context we consider.  Namely, we model 412,376 acts of gun violence recorded in the U.S.~between 2014 and 2024 (Figure \ref{fig:violence}).  

\begin{figure}[!t]
	\centering
	\includegraphics[width=0.45\textwidth]{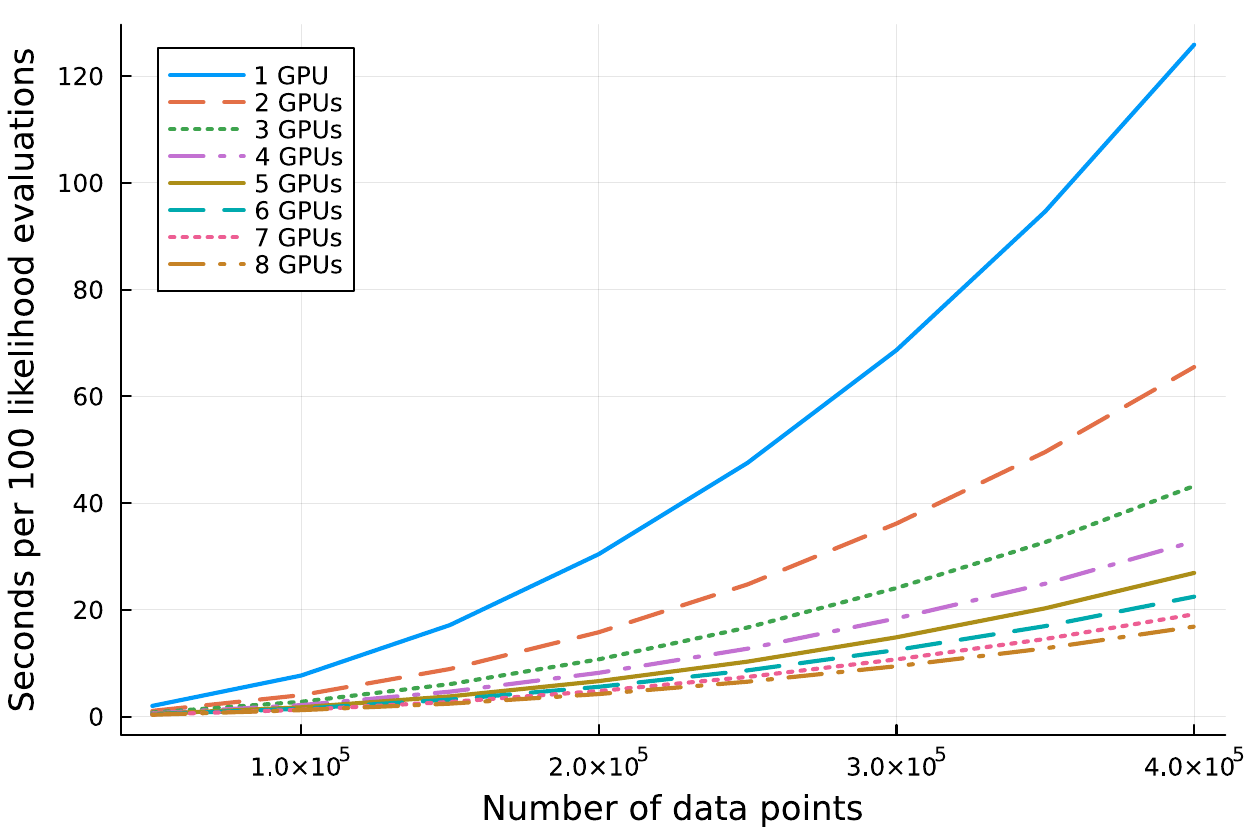}
		\includegraphics[width=0.45\textwidth]{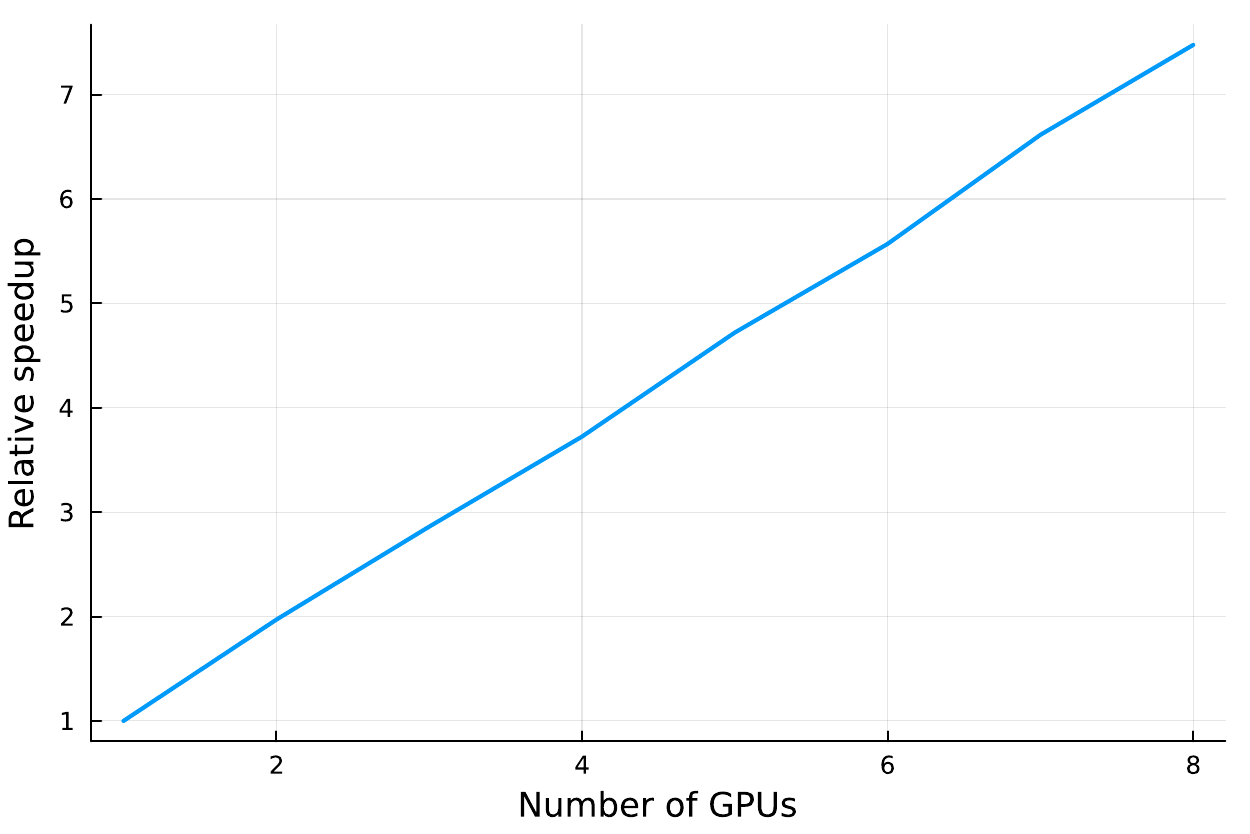}
	\caption{Absolute and relative speedups for spatiotemporal Hawkes process likelihood \eqref{eq:sthplik} with conditional intensity \eqref{eq:condInt} using different numbers of GPUs. We make the right plot using 412,376 data points, but trends hold for as many as 1,000,000. Notably, the single-GPU implementation is already over 200-times faster than an efficient single-core C++ implementation. Multi-GPU likelihood evaluations accelerate individual Metropolis-Hastings chains within our  many-GPU, many-chain Bayesian inference strategy.}\label{fig:perform}
\end{figure}

We fit our StHP model to this large-scale spatiotemporal data using the computational strategy for Bayesian inference advanced in \citet{ko2024scaling}.
As is the case for many of the likelihood-based methods discussed in Section \ref{sec:examp}, the likelihood associated with \eqref{eq:condInt} scales $O(N^2)$ in time-complexity.   Following \citet{ko2024scaling}, we use a multi-GPU extension of the single-GPU likelihood implementation of \citet{holbrook2021scalable}. Figure \ref{fig:perform} presents the relative performance of likelihood implementations using different numbers of GPUs; notably, the single-GPU likelihood implementation is already hundreds of times faster than an efficient C++ likelihood implementation and thousands of times faster than an analogous implementation in R \citep{holbrook2021scalable}.  We use four GPU nodes, each containing eight AMD Radeon Instinct MI50 GPUs, to generate 31 adaptive MH chains, with two GPUs assigned to each of the 16 chains run simultaneously.   Individual chains require 4.11 hours on average to complete, and the entire batch of chains less than 10 hours.  Across model parameters, we see a maximum $\hat{R}$ less than 1.003, minimum bulk- and tail-ESSs above 10,000 and 15,000, respectively, and minimum single-chain ESS above 300 \citep{rstan,coda}.

Table \ref{tab:medians} presents the empirical posterior medians and 95\% credible intervals (CrI) for each of the model parameters.  We are particularly interested in the posterior distributions of $\xi_0$, $\omega_x$ and $\omega_t$.  Because the temporal and spatial kernels used within the triggering kernel are normalized densities, the self-excitatory weight $\xi_0$ corresponds exactly to the branching number $\beta$ of \eqref{eq:branching}.  Thus, we conclude that the observed gun violence process is explosive: we expect each event to birth an additional 1.15 (95\% CrI: 1.15, 1.16) events.  The posterior median for the expected spatial distance between an event and its child event, conveyed by $\omega_x$, is 0.0205 degrees.  When translated to miles, the number varies depending on the latitude.  For example, 0.0205 degrees translates to 1.35 miles in Miami, FL, and 1.19 miles in Seattle, WA.  Finally, the posterior median for the expected time lapse between an event and its child event, conveyed by $\omega_t$, is 75.8 weeks (95\% CrI: 75.1, 76.5) or almost 1.5 years.  This number is orders-of-magnitude larger than the analogous number cited in \citet{holbrook2021scalable}, but that analysis involves gunshots recorded by AGLS and not the acts of gun \emph{violence} considered here.

\newcommand{\ra}[1]{\renewcommand{\arraystretch}{#1}}
\ra{1.2}
\begin{table}[!t]
	\centering
	\resizebox{1\textwidth}{!}{\begin{tabular}{lll} 
				\toprule
				Rate component& Parameter & Posterior median (95\% CrI)   \\
				\midrule
				Background  & Weight $\lambda_0$ & 0.0231 (0.0227, 0.0237)   \\
				& Temporal lengthscale $\tau_t$ (wks) & 2.62 (2.36, 2.93)   \\
				Self-excitatory & Weight $\xi_0$ & 1.15 (1.15, 1.16) \\
				&	Spatial lengthscale $\omega_x$ (deg$^*$)  & 0.0205 (0.0204, 0.0206)  \\
				& Temporal lengthscale $\omega_t$ (wks) & 75.8 (75.1, 76.5)  \\
				\bottomrule
		\end{tabular}}
		\caption{Posterior medians and 95\% credible intervals for parameters of a spatiotemporal Hawkes process model with conditional intensity \eqref{eq:condInt} applied to American gun violence data visualized in Figure \ref{fig:violence}.  The unit of measurement for the spatial lengthscale parameter $\omega_x$ is degrees (lattitude/longitude), for which the posterior median of 0.0205$^\circ$ converts to, e.g., 1.35 miles in Miami, FL, and 1.19 miles in Seattle, WA.  
		}\label{tab:medians}
	\end{table}

\section{Discussion}

   We have shown how to scale Bayesian inference to big data for a relatively simple StHP, but there is much work to be done in the advancement of computational tools for fitting HPs at scale.    In particular, there are major tensions between making HP models scalable and making them robust, flexible or expressive.  Furthermore, incorporating interesting structures underlying data, such as networks or multiple sources, often brings added computational complexity, thereby making fitting HP models adapted to these structures all the more difficult (e.g., Example \ref{ex:scott}).  We also see major tensions between scalability and user-friendliness: the large-scale analysis of gun violence of Section \ref{sec:meth} relies on hand-coded GPU kernels written in a language (OpenCL) that extends the C programming language.  NHPs such as that of Example \ref{ex:neural} are flexible and programmable using widely-used deep learning platforms such as PyTorch or TensorFlow, but their reliance on stochastic gradient descent for scalability means a lack of rigorous model identification and uncertainty quantification.  Nonetheless, we see NHPs and other deep generative models  as the dominant  modeling paradigm for event forecasting moving forward, with non-neural modeling reserved for interpretable and rigorous scientific inference that sheds light on natural mechanisms and, perhaps, informs policy.
   
   The ability to fit and compare a variety of competing HP models is crucial for meaningful scientific inference and currently lacking in big data contexts.  In Section \ref{sec:meth}, we fit the StHP with rate function \eqref{eq:condInt} to data consisting of over 400k  gun violence events, inferring scientifically interpretable parameters that shed light on the expanding nature of gun violence in the U.S.~(namely, we estimate that an average of 1.15 acts of violence arise from any individual act of violence) and the scales of its local spatial contagion.  While we remain enthusiastic about the ability of increasingly widespread high-performance computing technology to scale HP to big data, we cannot fully trust our inference, not having compared our model to other competitors.  Notably, we find the extremely narrow posterior credible intervals of Table \ref{tab:medians} particularly distressing.  Probably, the use of Bayesian nonparametric priors to define HP triggering kernels may help address this dearth of posterior uncertainty,  but one  needs to thoughtfully engage parallel computing resources for algorithmic developments that go hand in hand with model extensions.  The same statement applies for more prosaic modeling decisions such as that between exponential, power-law and Weibull temporal triggering kernels. 
   
  Many criteria for model comparison exist, but we believe that a model's forecasting accuracy is the only relevant criterion for big point process data.  This is because datasets consisting of many temporal events often become large when they span large swaths of time over which  stationarity assumptions fail.  Ideally, one would be able to train a candidate model on a nested sequence of data subsets, measuring different forecasting accuracy on subsequent data subsamples of varying time lengths. One might divide the gun violence data analyzed in Section \ref{sec:meth} into nested subsets  increasing at 1 month intervals, using HP model instances trained on each subset to generate probabilistic forecasts, say, 4 weeks into the future.  With 10 years of data, one would need to train each candidate model on roughly 120 subsets and use trained models to simulate large numbers of events.  It is for these reasons  we claim that proper scientific inference using HP applied to large sequences of events requires methods and computational tools for \emph{simulating} HP sample paths at scale, a subject we see nowhere in the literature.

\section*{Acknowledgments}

AJH is supported by grants NIH K25 AI153816, NSF DMS 2152774, and NSF DMS 2236854. We gratefully acknowledge generous support of Advanced Micro Devices, Inc., including the donation of parallel computing resources used for this research. We also thank the Gun Violence Archive for sharing its data. 

\bibliographystyle{chicago}
\bibliography{refs,refs2}

\end{document}